# $a$-axis YBa$_2$Cu$_3$O$_{7-x}$/PrBa$_2$Cu$_3$O$_{7-x}$/YBa$_2$Cu$_3$O$_{7-x}$

# trilayers with subnanometer rms roughness

Y. Eren Suyolcu,[1][†] Jiaxin Sun,[1][†] Berit H. Goodge,[2] Jisung Park,[1] Jürgen Schubert,[3] Lena F. Kourkoutis,[2,4] and Darrell G. Schlom[1,4,5]

[1]Department of Materials Sciences and Engineering, Cornell University, Ithaca, New York 14853, USA

[2]School of Applied and Engineering Physics, Cornell University, Ithaca, New York 14853, USA

[3]Peter Grünberg Institute (PGI-9) and JARA-Fundamentals of Future Information Technology, Forschungszentrum Jülich GmbH, 52425 Jülich, Germany

[4]Kavli Institute at Cornell for Nanoscale Science, Ithaca, New York 14853, USA

[5]Leibniz-Institut für Kristallzüchtung, Max-Born-Str. 2, 12489 Berlin, Germany

[†]These authors contributed equally to this work

**ABSTRACT**

We demonstrate $a$-axis YBa$_2$Cu$_3$O$_{7-x}$/PrBa$_2$Cu$_3$O$_{7-x}$/YBa$_2$Cu$_3$O$_{7-x}$ trilayers grown on (100) LaAlO$_3$ substrates with improved interface smoothness. The trilayers are synthesized by ozone-assisted molecular-beam epitaxy. The thickness of the PrBa$_2$Cu$_3$O$_{7-x}$ layer is held constant at 8 nm and the thickness of the YBa$_2$Cu$_3$O$_{7-x}$ layers is varied from 24 nm to 100 nm. X-ray diffraction measurements show all trilayers to have >95% $a$-axis content. The rms roughness of the thinnest trilayer is < 0.7 nm and this roughness increases with the thickness of the YBa$_2$Cu$_3$O$_{7-x}$ layers. The thickness of the YBa$_2$Cu$_3$O$_{7-x}$ layers also affects the transport properties: while all samples exhibit an onset of the superconducting transition at and above 85 K, the thinner samples show wider transition widths, $\Delta T_c$. High-resolution scanning transmission electron microscopy reveals coherent and chemically sharp interfaces, and that growth begins with a cubic (Y,Ba)CuO$_{3-x}$ perovskite phase that transforms into $a$-axis oriented YBa$_2$Cu$_3$O$_{7-x}$ as the substrate temperature is ramped up.

Shortly after the discovery of high-temperature superconductivity in $YBa_2Cu_3O_7$,[1,2] measurements showed that the superconducting proximity length along the $a$-axis ($\xi_a \approx 1.1$ nm)[3,4] of $YBa_2Cu_3O_{7-x}$ is nearly an order of magnitude longer than along the $c$-axis ($\xi_c \approx 0.1$ nm).[4,5] Note that $\xi_c$ is shorter than the distance between the $CuO_2$ planes. This difference makes the $a$-axis direction relevant to forming controlled and reproducible $YBa_2Cu_3O_{7-x}$ based Josephson junctions (JJs) for superconducting electronics.[6] Most JJs in $YBa_2Cu_3O_{7-x}$ have been made using epitaxial $YBa_2Cu_3O_{7-x}$ films oriented with the $c$-axis perpendicular to the film surface.[7] This is because such films, referred to as $c$-axis oriented films, exhibit the highest superconducting transition temperature ($T_c$), highest critical current density ($J_c$), and smoothest surface. JJs in such films are made in the (001) plane to exploit the longer in-plane coherence length, $\xi_a$. They occur at weak links present at grain boundaries[8–12] or formed by helium ion bombardment.[13] They are also made by introducing tunnel barriers by a ramp-junction process that involves patterning the $c$-axis $YBa_2Cu_3O_{7-x}$ film followed by epitaxial regrowth.[11,14,15]

A more direct approach to fabricate high quality JJs—one that involves pristine interfaces formed without breaking vacuum—is through the growth of $YBa_2Cu_3O_{7-x}$ films oriented with the $a$-axis perpendicular to the film surface (i.e., $a$-axis oriented films). This was recognized early on and numerous groups developed methods to grow $a$-axis oriented films[16,17] as well as JJs based upon them, e.g., $a$-axis oriented $YBa_2Cu_3O_{7-x}/PrBa_2Cu_3O_{7-x}/YBa_2Cu_3O_{7-x}$ trilayers.[6,18] Although JJs were successfully fabricated, the resulting junctions were neither controlled nor reproducible due to the significant roughness of the $a$-axis oriented $YBa_2Cu_3O_{7-x}$ films.[18] Note that the low-energy surface of $YBa_2Cu_3O_{7-x}$ is the (001) plane,[19] explaining the far smoother morphology of $c$-axis $YBa_2Cu_3O_{7-x}$ films compared to $a$-axis $YBa_2Cu_3O_{7-x}$ films.

To improve the quality of $a$-axis YBa$_2$Cu$_3$O$_{7-x}$ films and heterostructures, several techniques have been employed when using PrBa$_2$Cu$_3$O$_{7-x}$ as either a buffer layer or a barrier layer.[20] These include increasing the substrate temperature, either gradually[21] or with a step-like ramp after the $a$-axis buffer layer has nucleated,[6,20,22,23] and even performing this in tandem with ramping down the background oxidant gas pressure.[21,24] Nonetheless, interface roughness has remained a major challenge for samples showing good superconducting transitions (e.g., an rms of ~10 nm for $a$-axis YBa$_2$Cu$_3$O$_{7-x}$ grown on (100) LaAlO$_3$)[21] as has avoiding the unwanted nucleation of $c$-axis YBa$_2$Cu$_3$O$_{7-x}$ or PrBa$_2$Cu$_3$O$_{7-x}$ when the temperature is ramped.[22,23]

Following the initial pioneering studies on $a$-axis YBa$_2$Cu$_3$O$_{7-x}$/PrBa$_2$Cu$_3$O$_{7-x}$/YBa$_2$Cu$_3$O$_{7-x}$ trilayers and a recognition of the challenges involved in making a viable JJ technology by this approach, work on this system has all but ceased. Now decades later, we revisit this challenge harnessing the improvements that have been made in the intervening years in thin film growth methods. Using ozone-assisted molecular-beam epitaxy (MBE), we grow $a$-axis YBa$_2$Cu$_3$O$_{7-x}$/PrBa$_2$Cu$_3$O$_{7-x}$/YBa$_2$Cu$_3$O$_{7-x}$ trilayers paying particular attention to growth conditions that yield smooth surfaces. We study the thickness dependence of the surface roughness as well as the superconducting transition width. Our results, including cross-sectional scanning transmission electron microscopy with electron energy loss spectroscopy (STEM-EELS) to characterize the interfaces with chemical specificity, demonstrate that the interface roughness can be decreased significantly to a level comparable to the thickness of relevant tunneling barrier layers. The substantial improvement in interface smoothness that we observe in $a$-axis YBa$_2$Cu$_3$O$_{7-x}$/PrBa$_2$Cu$_3$O$_{7-x}$/YBa$_2$Cu$_3$O$_{7-x}$ trilayers suggests that $a$-axis YBa$_2$Cu$_3$O$_{7-x}$-based JJs with requisite smoothness to provide the precise thickness control of the tunnel barrier needed for a JJ technology is achievable.

$YBa_2Cu_3O_{7-x}/PrBa_2Cu_3O_{7-x}/YBa_2Cu_3O_{7-x}$ trilayers with 24 nm, 32 nm, 64 nm, and 100 nm thick $YBa_2Cu_3O_{7-x}$ layers, in which the $PrBa_2Cu_3O_{7-x}$ layer thickness is kept constant at 8 nm, were grown on 10 mm × 10 mm (100)-oriented $LaAlO_3$ substrates by ozone-assisted MBE (Fig. 1(a)). Although high quality $a$-axis $YBa_2Cu_3O_{7-x}$ films have been grown on (100) $LaSrGaO_4$ substrates,[25] we used (100) $LaAlO_3$ substrates in this work because our goal is to identify a path that can be scaled to large diameters to enable its translation to a viable technology. 3-inch diameter $LaAlO_3$ substrates are currently available; in the past, even 4-inch diameter $LaAlO_3$ substrates were commercially produced.[26]

The $YBa_2Cu_3O_{7-x}/PrBa_2Cu_3O_{7-x}/YBa_2Cu_3O_{7-x}$ trilayers were synthesized in a Veeco GEN10 MBE. Yttrium (99.6%), barium (99.99%), praseodymium (99.1%), and copper (99.99%) were evaporated from thermal effusion cells with fluxes of $1.1\times10^{13}$ cm$^{-2}$s$^{-1}$, $2.2\times10^{13}$ cm$^{-2}$s$^{-1}$, and $3.3\times10^{13}$ cm$^{-2}$s$^{-1}$, respectively. Prior to growth, the (100) $LaAlO_3$ substrates (CrysTec GmbH) were etched in boiling water, annealed at 1300 °C in air for 10 hours, and then etched again in boiling water, to obtain an $AlO_2$-terminated surface with a step-and-terrace morphology.[27] Following this surface treatment, the backside of the (100) $LaAlO_3$ substrates were coated with a 10 nm thick titanium adhesion layer followed by 200 nm of platinum, enabling the otherwise transparent substrates to be radiatively heated during MBE growth. The $YBa_2Cu_3O_{7-x}$ (or $PrBa_2Cu_3O_{7-x}$) layers were grown by simultaneously depositing yttrium (or praseodymium), barium, and copper onto the heated substrate under a continuous flux of distilled ozone (~80% $O_3$ + 20% $O_2$) yielding a background pressure of $1\times10^{-6}$ Torr. After growth, the samples were cooled to under 100 °C in the same pressure of distilled ozone in which they were grown before turning off the ozone molecular beam and removing the samples from vacuum.

Because $YBa_2Cu_3O_{7-x}$ is a point compound that is unable to accommodate appreciable off-stoichiometry,[28] flux calibration presents a significant challenge where secondary impurity phases nucleate easily and significantly degrade film quality.[29] We tackle this challenge by separately calibrating the flux of each element by growing binary oxides of the constituents, namely $Y_2O_3$, $PrO_2$, BaO, and CuO. From these separate binary flux calibrations, the temperatures of the effusion cells containing yttrium, barium, praseodymium, and copper are adjusted to match the desired 1:2:3 flux ratio among Y(Pr):Ba:Cu. The temperature of the substrate is measured during growth by a thermocouple ($T_{Tc}$) that is positioned close to, but not in direct contact with the substrate, and by an optical pyrometer ($T_{Pyr}$) operating at a wavelength of 1550 nm. The growth of the trilayers starts at low-temperature, $T_{Tc} \approx 420\ °C$ ($T_{Pyr} \approx 530\ °C$), resulting in a cubic perovskite $(Y,Ba)CuO_{3-x}$ phase[30] for the first few layers and ends at $T_{Tc} \approx 570\ °C$ ($T_{Pyr} \approx 620\ °C$) following a temperature-ramping procedure. Details of the flux calibration method (including the characterization of individual binary oxides) are presented in Figs. S1-S5 of the Supplementary Material. Also shown are the temperature-ramping details and the *in-situ* reflection high-energy electron diffraction (RHEED) characterization of a reference *a*-axis $YBa_2Cu_3O_{7-x}$ single-phase film grown as part of the optimization of the growth procedure (Fig. S6).

During growth the films were monitored by *in-situ* RHEED with KSA-400 software and a Staib electron gun operating at 13 kV and 1.45 A. RHEED images taken during the growth of the 24 nm $YBa_2Cu_3O_{7-x}$/8 nm $PrBa_2Cu_3O_{7-x}$/24 nm $YBa_2Cu_3O_{7-x}$ trilayer are shown in Figs. 1(b)-1(g). The structural quality and the *a*-axis/*c*-axis ratio of the samples was explored using a PANalytical

Empyrean x-ray diffractometer (XRD) at 45 kV and 40 mA with Cu $K\alpha_1$ radiation (1.54057 Å). For surface morphological characterization of the films, *ex situ* atomic force microscopy (AFM) measurements were conducted using an Asylum Cypher ES Environmental AFM system. Resistance as a function of temperature measurements were carried out using a homemade four-point van der Pauw geometry system that slowly dips the samples into a Dewar of liquid helium.

Detailed investigations of the films were conducted using atomic-resolution scanning transmission electron microscopy (STEM). Cross-sectional TEM specimens were prepared by focused ion beam (FIB) lift-out with a Thermo Fisher Helios G4 UX Dual Beam system. The samples were imaged on an aberration-corrected FEI Titan Themis at 300 kV. STEM high-angle annular dark-field (HAADF) imaging was performed with a probe convergence semi-angle of 21.4 mrad and inner and outer collection angles from 68-340 mrad. STEM electron energy loss spectroscopy (EELS) measurements were performed on the same Titan system equipped with a 965 GIF Quantum ER and Gatan K2 Summit direct detector operated in electron counting mode, with a beam current of ~50 pA and scan times of 2.5 or 5 ms per 0.4 Å pixel. A multivariate weighted principal component analysis routine (MSA Plugin in Digital Micrograph) is used to decrease the noise level in STEM data.[31]

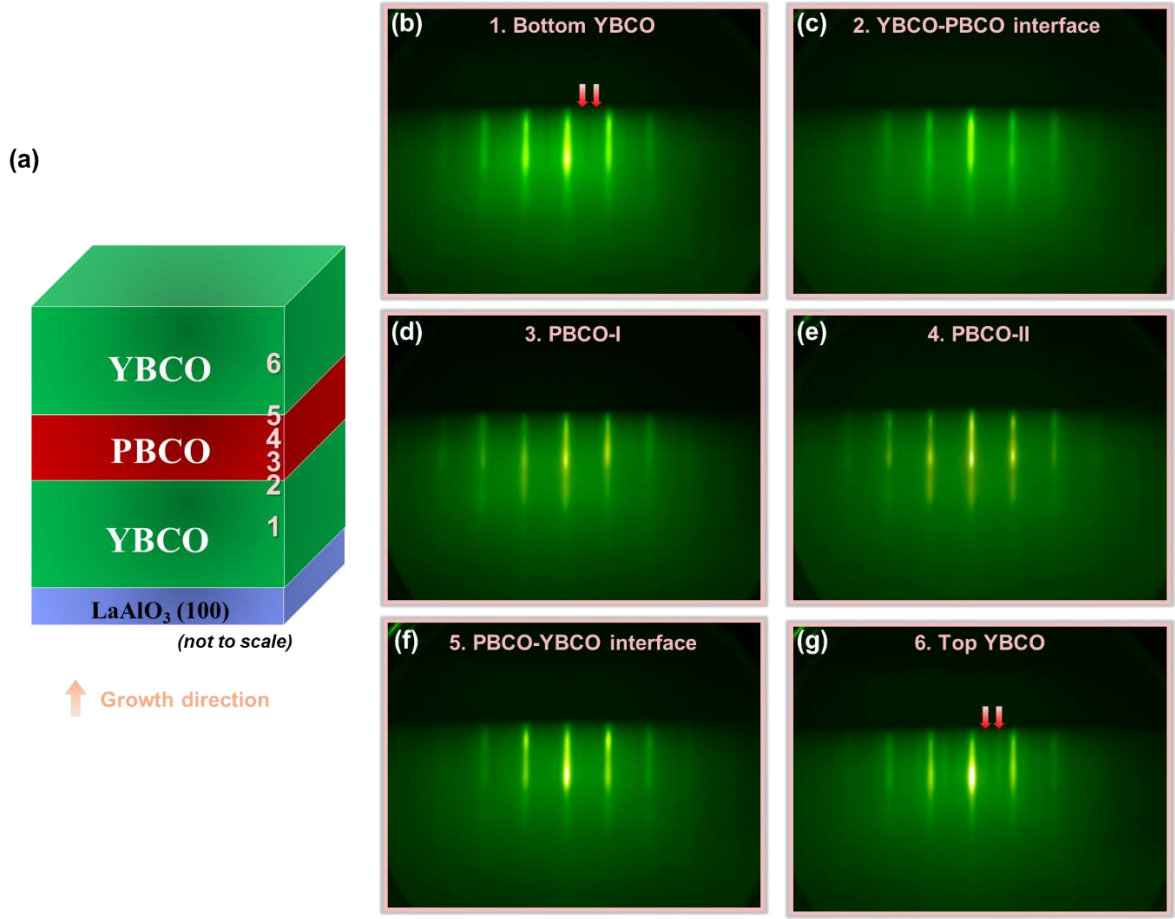

**Figure 1.** (a) Schematic of the YBa$_2$Cu$_3$O$_{7-x}$/PrBa$_2$Cu$_3$O$_{7-x}$/YBa$_2$Cu$_3$O$_{7-x}$ trilayers grown on (100) LaAlO$_3$ substrates. (b-g) Real-time RHEED images acquired along the [010] azimuth of the (100) LaAlO$_3$ substrate during the growth of the 24 nm YBa$_2$Cu$_3$O$_{7-x}$/8 nm PrBa$_2$Cu$_3$O$_{7-x}$/24 nm YBa$_2$Cu$_3$O$_{7-x}$ trilayer at the six different times schematically illustrated in (a). The RHEED patterns are from the bottom YBa$_2$Cu$_3$O$_{7-x}$ layer (b), YBa$_2$Cu$_3$O$_{7-x}$/PrBa$_2$Cu$_3$O$_{7-x}$ interface (c), PrBa$_2$Cu$_3$O$_{7-x}$ layer (d,e), PrBa$_2$Cu$_3$O$_{7-x}$/YBa$_2$Cu$_3$O$_{7-x}$ interface, and (f) top YBa$_2$Cu$_3$O$_{7-x}$ layer. The red arrows added to (b) and (g) point to the diffraction streaks associated with the *c*-axis of the YBa$_2$Cu$_3$O$_{7-x}$ lying in-plane.

The structural quality of the samples is assessed by XRD measurements. In the coupled $\theta$-$2\theta$ XRD scans in Fig. 2(a), only $h00$, $0k0$, and $00\ell$ reflections of the YBa$_2$Cu$_3$O$_{7-x}$ and PrBa$_2$Cu$_3$O$_{7-x}$ phases are indexed, indicating that the film only contains phases with the desired stoichiometry; they are free of impurity phases associated with off-stoichiometry. With increasing $YBa_2Cu_3O_{7-x}$ layer thicknesses, $00\ell$ reflections emerge showing the nucleation and propagation of $c$-axis grains in the films. Off-axis $\phi$ scans of the 102 family of reflections of the orthorhombic $YBa_2Cu_3O_{7-x}$ / $PrBa_2Cu_3O_{7-x}$ at $\chi \approx 56.6°$ and $\chi \approx 33.4°$ are used to measure the $a$-axis and $c$-axis content of the orthorhombic grains, respectively. Note that $\chi = 90$ aligns the diffraction vector to be perpendicular to the plane of the substrate.[32] In the 102 $\phi$ scan of the trilayer sample shown in Fig. 2(b), four peaks associated with the $a$-axis grains are observed corresponding to 90° in-plane rotational twinning: the $c$-axis of the $YBa_2Cu_3O_{7-x}$ and $PrBa_2Cu_3O_{7-x}$ is aligned parallel to the [010] direction of the (100) $LaAlO_3$ substrate in one set of twin domains and parallel to the [001] direction of the (100) $LaAlO_3$ substrate in the other set of twin domains.[17,20,33,34] No intensity associated with $c$-axis grains is observed indicating that the film contains no $c$-axis grains within the resolution of our XRD scan. The off-axis $\phi$ scans of all trilayer samples shown in Fig. S5 indicate that all four trilayers have more than 95% $a$-axis content in the $Y(Pr)Ba_2Cu_3O_{7-x}$ orthorhombic phase. In addition to the orthorhombic phases, we also observe a cubic perovskite phase. This phase has been previously reported in the literature as a low-temperature, kinetically-stabilized $I$-centered cubic phase[35] or primitive simple-cubic phase.[36] The formation of this phase and its role in stabilizing the $a$-axis $YBa_2Cu_3O_{7-x}/PrBa_2Cu_3O_{7-x}/YBa_2Cu_3O_{7-x}$ trilayers is discussed below in tandem with its observation by HAADF-STEM. In the reciprocal space map (RSM) around the $LaAlO_3$ $\bar{1}03$ reflection (pseudocubic) in Fig. 2(c), we also observe a perovskite-like $\bar{1}03$ reflection (denoted $p$-$(Y,Ba)CuO_{3-x}$) and the orthorhombic phase $30\bar{3}/3\bar{1}0$ and $033/\bar{1}30$ reflections associated with the $a$-axis and $b$-axis $YBa_2Cu_3O_{7-x}$ grains, respectively..

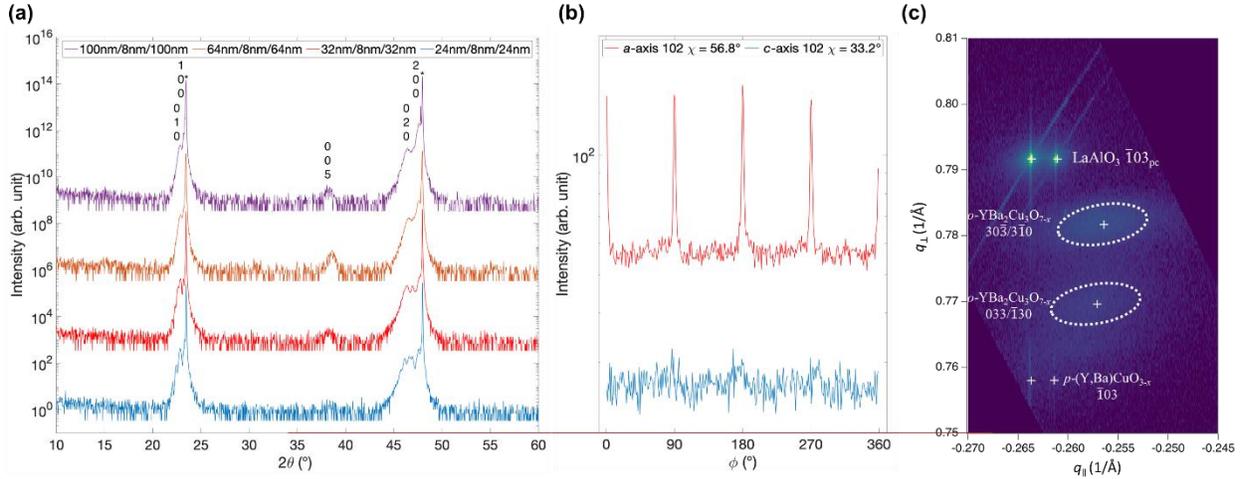

**Figure 2.** X-ray diffraction of YBa$_2$Cu$_3$O$_{7-x}$/PrBa$_2$Cu$_3$O$_{7-x}$/YBa$_2$Cu$_3$O$_{7-x}$ trilayers with 24 nm, 32 nm, 64 nm, and 100 nm thick YBa$_2$Cu$_3$O$_{7-x}$ layers. (a) $\theta$-$2\theta$ scans of the trilayers show only $h00$, $0k0$, and $00\ell$ reflections. (b) Off-axis 102 reflection $\phi$ scans at $\chi \approx 56.6°$ (red) and $\chi \approx 33.4°$ (blue) of the trilayer with 24 nm thick YBa$_2$Cu$_3$O$_{7-x}$ layers showing the absence of $c$-axis grains. (c) RSM around the LaAlO$_3$ $\bar{1}03$ reflection (pseudocubic) of this same trilayer with 24 nm thick YBa$_2$Cu$_3$O$_{7-x}$ layers showing the $a$-axis and $b$-axis orthorhombic $30\bar{3}/3\bar{1}0$ and $033/\bar{1}30$ reflections of YBa$_2$Cu$_3$O$_{7-x}$, as well as the perovskite (Y,Ba)CuO$_{3-x}$ $\bar{1}03$ reflection. The positions of these film reflections are indicated by the "+" symbols and dashed ellipses near the corresponding reflection labels.

The surface morphologies of the same as-grown YBa$_2$Cu$_3$O$_{7-x}$/PrBa$_2$Cu$_3$O$_{7-x}$/YBa$_2$Cu$_3$O$_{7-x}$ trilayers were established by *ex situ* AFM in tapping mode. With increasing YBa$_2$Cu$_3$O$_{7-x}$ layer thickness, the elongated YBa$_2$Cu$_3$O$_{7-x}$ grains as well as the in-plane 90° rotational twinning of these rectangular-shaped features become visible in the 2 µm × 2 µm topography scans presented in Figs. 3(a)-3(d). This morphology arises from the much slower growth rate of YBa$_2$Cu$_3$O$_{7-x}$ grains along [001] than in the (001) plane.[37] The root-mean-square (rms) roughness also increases with increasing YBa$_2$Cu$_3$O$_{7-x}$ layer thickness from 0.62 nm in the thinnest 24 nm/8 nm/24 nm trilayer to 2.3 nm in the thickest 100 nm/8 nm/100 nm trilayer. Surface roughness is an important metric affecting the yield and electrical performance of YBa$_2$Cu$_3$O$_{7-x}$-based JJs involving extrinsic interfaces, i.e., tunnel barriers. The 0.62 nm rms roughness we observe is the smoothest reported in the literature and a significant reduction from the 11.3 nm measured previously on *a*-axis YBa$_2$Cu$_3$O$_{7-x}$/PrBa$_2$Cu$_3$O$_{7-x}$ bilayers with 270 nm thick YBa$_2$Cu$_3$O$_{7-x}$ layers grown on (100) LaAlO$_3$ substrates.[21]

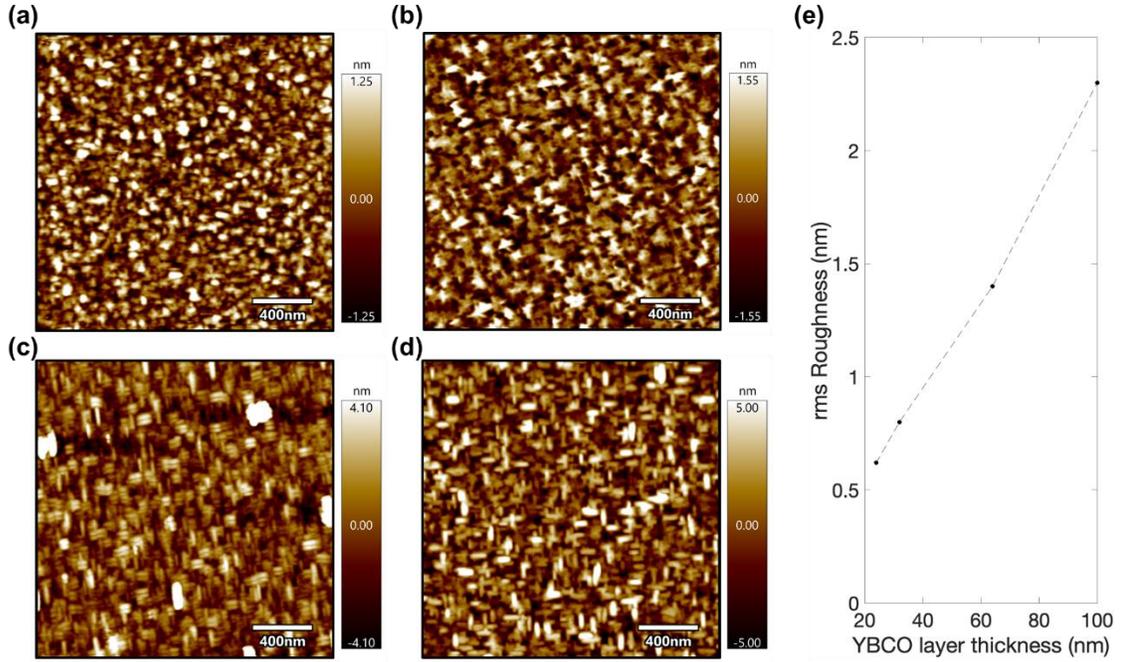

**Figure 3.** Surface morphology of the YBa$_2$Cu$_3$O$_{7-x}$/PrBa$_2$Cu$_3$O$_{7-x}$/YBa$_2$Cu$_3$O$_{7-x}$ trilayers revealed by AFM. (a)-(d) 2 µm × 2 µm topography scans of the 24 nm/8 nm/24 nm, 32 nm/8 nm/32 nm, 64 nm/8 nm/64 nm, and 100 nm/8 nm/100 nm trilayers in tapping mode, respectively. (e) rms roughness calculated from (a)-(d) as a function of the YBa$_2$Cu$_3$O$_{7-x}$ layer thickness. The dotted line serves as a guide for the eye.

The resistance as a function of temperature (*R-T*) was measured on the same YBa$_2$Cu$_3$O$_{7-x}$/PrBa$_2$Cu$_3$O$_{7-x}$/YBa$_2$Cu$_3$O$_{7-x}$ trilayers; the results are presented in Fig. 4. As is evident from the *R-T* plots in Fig. 4(a), all trilayers superconduct. The normal state resistance decreases and the onset temperature of the superconducting transition (*T*$_{onset}$) increases with increasing YBa$_2$Cu$_3$O$_{7-x}$ layer thickness—from 85 K for the 24nm/8nm/24nm trilayer to 90 K for the 100nm/8nm/100nm trilayer, as shown in Fig. 4(b). We define $T_{\text{onset}}$ as the temperature at which the resistance falls below a linear extrapolation of the $R$ vs. $T$ behavior from its slope in the 200-300 K regime. The superconducting transition width ($\Delta T_c$), here defined as the temperature difference between $T_{\text{onset}}$ and the temperature at which the resistance is zero (within the noise of our measurement), $\Delta T_c$, decreases with increasing YBa$_2$Cu$_3$O$_{7\text{-}x}$ layer thickness from 29 K for the 24 nm/8 nm/24 nm trilayer to 10 K for the 100 nm/8 nm/100 nm trilayer, as seen in Fig. 4(c). Compared to $c$-axis YBa$_2$Cu$_3$O$_{7\text{-}x}$ films, however, these transition widths are still relatively broad.[38] Such behavior is ubiquitous in twinned $a$-axis YBa$_2$Cu$_3$O$_{7\text{-}x}$ films[17,20,21,24] especially when the thickness of the $a$-axis YBa$_2$Cu$_3$O$_{7\text{-}x}$ is under 100 nm.[18,39] It may arise from local disorder and inhomogeneities in the samples, insufficient oxidation, or from the degradation of the samples over time.[40]

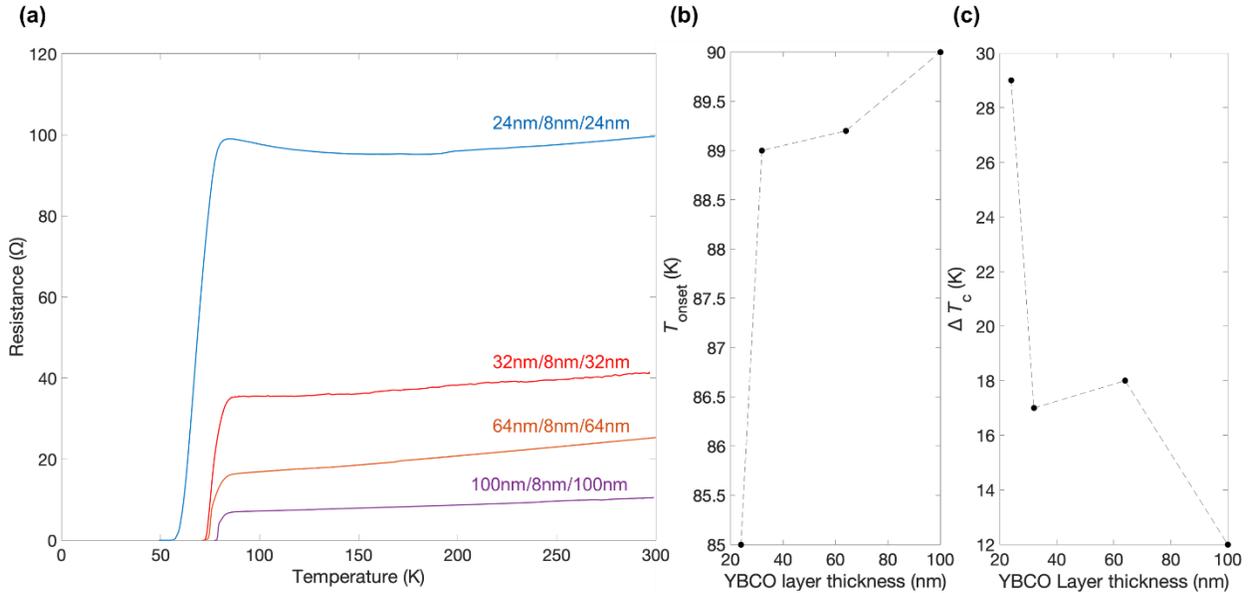

**Figure 4.** Transport properties of the same YBa$_2$Cu$_3$O$_{7\text{-}x}$/PrBa$_2$Cu$_3$O$_{7\text{-}x}$/YBa$_2$Cu$_3$O$_{7\text{-}x}$ trilayers. (a) Resistance as a function of temperature, (b) onset of the superconducting transition ($T_{\text{onset}}$) as a function of YBa$_2$Cu$_3$O$_{7\text{-}x}$ layer thickness, and (c) superconducting transition width ($\Delta T_c$) as a function of YBa$_2$Cu$_3$O$_{7\text{-}x}$ layer thickness. The dotted lines are guides to the eye.

To reveal the microstructure and interface abruptness of the samples, we studied two trilayer samples with cross-sectional high-resolution STEM. A low-magnification HAADF-STEM image of the 24 nm/8 nm/24 nm $YBa_2Cu_3O_{7-x}$/$PrBa_2Cu_3O_{7-x}$/$YBa_2Cu_3O_{7-x}$ trilayer shown in Fig. 5(a) is representative of the complete sample. Individual layers are distinguished as darker and brighter regions due to the atomic number ($Z$) contrast[41] of HAADF imaging. The $PrBa_2Cu_3O_{7-x}$ layer gives brighter contrast compared to the $YBa_2Cu_3O_{7-x}$ layer because praseodymium ($Z_{Pr} = 59$) is heavier than yttrium ($Z_Y = 39$). The $LaAlO_3$ substrate also shows relatively bright contrast for the same reason ($Z_{La} = 57$). A higher magnification image (Fig. 5(b)) focusing on a representative interface region reveals that the interfaces in the $YBa_2Cu_3O_{7-x}$/$PrBa_2Cu_3O_{7-x}$/$YBa_2Cu_3O_{7-x}$ trilayer are coherent. In neither low-magnification nor in high-magnification scans were $c$-axis grains observed in our STEM images, consistent with the high volume fraction of $a$-axis growth measured by XRD. Nevertheless, structural coherence does not prove chemical abruptness at interfaces involving cuprate high-temperature superconductors.[42,43]

The chemical abruptness of the $YBa_2Cu_3O_{7-x}$/$PrBa_2Cu_3O_{7-x}$/$YBa_2Cu_3O_{7-x}$ interfaces was assessed by atomic-resolution elemental mapping via STEM-EELS. Figures 5(c)-5(e) show the elemental maps obtained using Pr–$M_{5,4}$ (red), Y–$L_{3,2}$ (green), and Ba–$M_{5,4}$ (blue) edges in the region outlined by the tan rectangle in Fig. 5(a). A red, green, blue (RGB) overlay of the elemental maps from this region is shown in Fig. 5(f), while Fig. 5(g) shows the simultaneously acquired ADF-STEM image of the same region. Atomic-resolution EELS maps reveal abrupt interface profiles, corroborating the STEM-HAADF images. Both interfaces show minimal Y-Pr intermixing, although some asymmetry of the interface profiles is seen. The lower $YBa_2Cu_3O_{7-x}$/$PrBa_2Cu_3O_{7-x}$ interface shows a nearly perfect interface profile free of Y–Pr intermixing; the upper interface ($PrBa_2Cu_3O_{7-x}/YBa_2Cu_3O_{7-x}$) presents a slightly rougher local profile with a roughness limited to 1-2 monolayers.

The roughness of the interfaces revealed by STEM and STEM-EELS in Fig. 5 is consistent with the qualitative observations made during growth by *in situ* RHEED (Figs. 1(b)-1(g)) of this same $YBa_2Cu_3O_{7-x}/PrBa_2Cu_3O_{7-x}/YBa_2Cu_3O_{7-x}$ trilayer. The arrowed streaks of *a*-axis oriented $YBa_2Cu_3O_{7-x}$ in Fig. 1(b) promptly disappear in transitioning from the lower $YBa_2Cu_3O_{7-x}$ layer to the $PrBa_2Cu_3O_{7-x}$ barrier layer in Fig. 1(c), indicating that the $PrBa_2Cu_3O_{7-x}$ barrier layer uniformly covers the lower $YBa_2Cu_3O_{7-x}$ layer. At the upper interface, however, it takes noticeably longer for the arrowed streaks of the upper $YBa_2Cu_3O_{7-x}$ layer to reappear (Figs. 1(f) and 1(g)). Further, the time that it takes for the arrowed streaks of *a*-axis oriented $YBa_2Cu_3O_{7-x}$ to reappear in going from the $PrBa_2Cu_3O_{7-x}$ barrier to the $YBa_2Cu_3O_{7-x}$ upper layer takes progressively longer for the thicker trilayers. This is consistent with the increased surface roughness seen by AFM in Fig. 3 as the thickness of the $YBa_2Cu_3O_{7-x}$ layers increases.

In addition to the coherent and chemically sharp interfaces, some defects were observed by STEM. For example, intergrowths of an extra Cu-O layer intercalated into the $YBa_2Cu_3O_{7-x}$ structure to locally form $YBa_2Cu_4O_{8-x}$ (Fig. S7) are seen. Such intergrown layers are well-known and common in $YBa_2Cu_3O_{7-x}$—in bulk, thin-films, and heterostructures.[44–46]

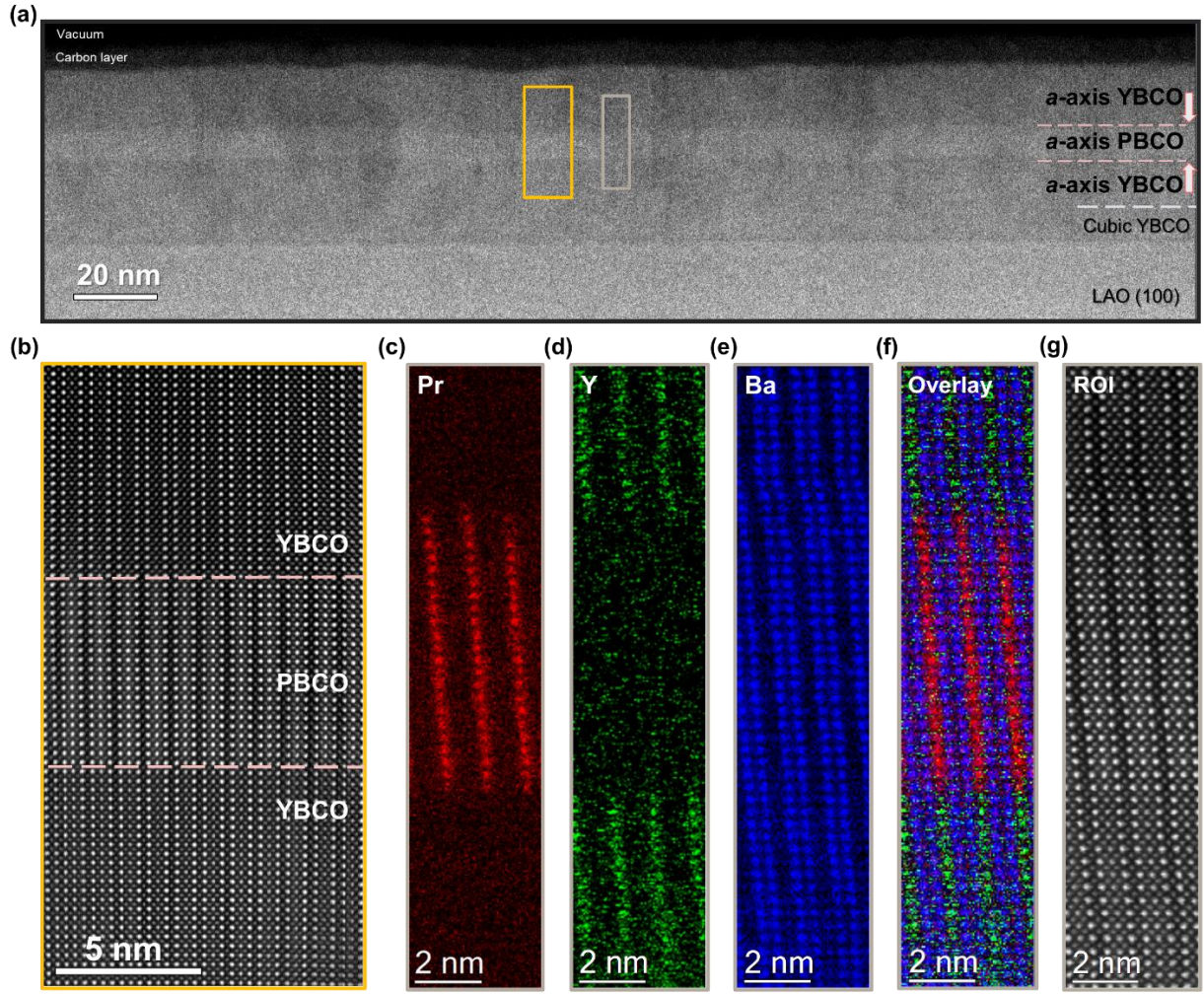

**Figure 5.** (a) Low-magnification cross-sectional HAADF-STEM image of the 24 nm/8 nm/24 nm $YBa_2Cu_3O_{7-x}$/$PrBa_2Cu_3O_{7-x}$/$YBa_2Cu_3O_{7-x}$ trilayer revealing the microstructure and interface abruptness at the atomic scale. Individual $YBa_2Cu_3O_{7-x}$ and $PrBa_2Cu_3O_{7-x}$ layers are separated using dashed lines and the pink arrows indicate the interfaces. (b) High-magnification scan of the area highlighted by the orange rectangle in (a) demonstrates the interfaces are fully coherent. (c)–(e) Atomically resolved Pr–$M_{5,4}$ edge (red), Y–$L_{3,2}$ edge (green), and Ba–$M_{5,4}$ edge (blue) elemental maps evidencing the sharp chemical abruptness of the interfaces. (f) The RGB overlay and (g) the simultaneously acquired ADF-STEM image of the same region, outlined by the tan rectangle in (a).

The cross-sectional HAADF-STEM imaging also unveils the location of the cubic perovskite $(Y,Ba)CuO_{3-x}$ phase detected in the XRD measurements. The thickness of the cubic $(Y,Ba)CuO_{3-x}$ layer is found to be ~10 nm and it is located under the bottom $YBa_2Cu_3O_{7-x}$ layer (Fig. S7(a)). This cubic $(Y,Ba)CuO_{3-x}$ layer forms at the start of growth when the substrate is coldest and surface diffusion is most constrained. Yttrium and barium are unable to diffuse sufficiently far to establish the Y-Ba-Ba-… ordered arrangement found in the unit cell of $YBa_2Cu_3O_{7-x}$; instead yttrium and barium share the $A$-site of the resulting perovskite structure, with copper on the $B$-site.[47]

As the temperature of the substrate is ramped, the diffusion lengths increase, and in-plane structural order emerges. The resulting $a$-axis $YBa_2Cu_3O_{7-x}$ grains grow epitaxially in one of two symmetry equivalent orientations: with the $c$-axis parallel to either [010] or [001] of the cubic $(Y,Ba)CuO_{3-x}$ layer on which they nucleate on the (100) $LaAlO_3$ substrate. One set of such domains is clearly seen in Fig. S7: the set with the $c$-axis along the horizontal direction of the image. The other set, with the $c$-axis oriented into the plane of the image, are more difficult to establish because their spacing along the horizontal direction is the same perovskite spacing as the cubic $(Y,Ba)CuO_{3-x}$ layer on which they nucleated.

Our hypothesis is that the ~10 nm thick cubic $(Y,Ba)CuO_{3-x}$ layer only lies under the $a$-axis oriented $YBa_2Cu_3O_{7-x}$ layer and that the regions in which this perovskite structure appears to extend further, i.e., through and all the way to the surface of the trilayer, are actually the set of $a$-axis domains oriented with the $c$-axis running into the plane of the image. This hypothesis is consistent with the grain size of the $a$-domains seen in the AFM images (Figs. 3(a)-3(d)) as well as published by others for $a$-axis $YBa_2Cu_3O_{7-x}$ grown on (100) $LaAlO_3$.[17,20,33,34,47,48] We know from the XRD $\phi$-scans (Figs. 2(b) and S5) that there is an equal volume fraction of both 90° in-plane rotation twin variants and although the volume sampled in our STEM investigation is small, this hypothesis is also consistent with our STEM observations. Once the substrate temperature is sufficiently high that the $a$-axis

YBa$_2$Cu$_3$O$_{7-x}$ grains nucleate, both twin variants continue through the entire YBa$_2$Cu$_3$O$_{7-x}$/PrBa$_2$Cu$_3$O$_{7-x}$/YBa$_2$Cu$_3$O$_{7-x}$ trilayer.

Lastly, in order to gain insights on the effect of *c*-axis grains in the trilayers, we perform additional cross-sectional STEM investigations on a less-ideal 32 nm/8 nm/32 nm sample. XRD shows the sample chosen to contain a higher volume fraction (16%) of *c*-axis oriented YBa$_2$Cu$_3$O$_{7-x}$/PrBa$_2$Cu$_3$O$_{7-x}$ (Fig. S8) and to have a higher rms roughness than the 32 nm/8 nm/32 nm trilayer characterized in Figs. 2-4. HAADF-STEM imaging (Fig. S9) of this less-ideal 32 nm/8 nm/32 nm trilayer confirms the presence of *c*-axis oriented grains in the structure and also demonstrates the rougher interfaces. Although the interfaces are rougher, STEM-EELS (Fig. S10) shows that they remain chemically abrupt. These results, when evaluated together, explain the rougher surfaces of the thicker samples. The *c*-axis grain formation in the bottom YBa$_2$Cu$_3$O$_{7-x}$ layer not only disturbs the PrBa$_2$Cu$_3$O$_{7-x}$ layer (and interface) profiles, but also directly influences the top surface roughness with changes in the local structural homogeneity in the first layers of the growth. The strong correlation between surface roughness and the volume fraction of *c*-axis grains in *a*-axis YBa$_2$Cu$_3$O$_{7-x}$ films has been previously noted.[39] To avoid *c*-axis oriented YBa$_2$Cu$_3$O$_{7-x}$, we initiate growth at a substrate temperature where only cubic (Y,Ba)CuO$_{3-x}$ can nucleate.

In conclusion, we revisited the growth of *a*-axis YBa$_2$Cu$_3$O$_{7-x}$/PrBa$_2$Cu$_3$O$_{7-x}$/YBa$_2$Cu$_3$O$_{7-x}$ trilayers and were able to improve their structural quality. By leveraging a temperature-ramping procedure that begins with a cubic (Y,Ba)CuO$_{3-x}$ buffer layer, we have grown high-quality *a*-axis trilayers as confirmed by *ex-situ* XRD measurements. AFM investigations revealed the improved surface quality with rms roughness that is less than $\zeta_a$ for the thinnest YBa$_2$Cu$_3$O$_{7-x}$/PrBa$_2$Cu$_3$O$_{7-x}$/YBa$_2$Cu$_3$O$_{7-x}$ trilayers. STEM analyses unveil the interrelation between $c$-axis oriented regions and surface roughness. Resistivity vs. temperature measurements exhibit an onset of the superconducting transition at $T_{onset} \sim 85$ K and also the widening of the superconducting transition width with decreasing $YBa_2Cu_3O_{7-x}$ film thickness. Sharp and coherent interfaces with limited elemental intermixing are evidenced by atomic-resolution HAADF-STEM and STEM-EELS. Our findings suggest that with precise control of the growth conditions, the sharp interfaces and smooth surfaces required in $a$-axis-based $YBa_2Cu_3O_{7-x}$ heterostructures for high-performance Josephson junctions and other oxide electronics are within reach.

## ACKNOWLEDGMENTS

This work was primarily supported by Ambature, Inc. B.H.G. and L.F.K. acknowledge support by the Department of Defense Air Force Office of Scientific Research (No. FA 9550-16-1-0305). The authors thank Ronald Kelly, Michael Lebby, Davis Hartman, Mitch Robson, and Ivan Bozovic for fruitful discussions. This work made use of a Helios FIB supported by NSF (DMR-1539918) and the Cornell Center for Materials Research (CCMR) Shared Facilities, which are supported through the NSF MRSEC Program (Grant No. DMR-1719875). The authors acknowledge Malcolm Thomas, Donald Werder, John Grazul, and Mariena Silvestry Ramos for assistance in the Electron Microscopy CCMR facilities. The FEI Titan Themis 300 was acquired through Grant No. NSF-MRI-1429155, with additional support from Cornell University, the Weill Institute, and the Kavli Institute at Cornell. This work also made use of the CESI Shared Facilities partly sponsored by the NSF (Grant No. DMR-1338010) and the Kavli Institute at Cornell. Substrate preparation was performed in part at the Cornell NanoScale Facility, a member of the National Nanotechnology Coordinated Infrastructure (NNCI), which is supported by the NSF (Grant No. NNCI-2025233). The authors thank Sean C. Palmer for his assistance with substrate preparation.

## Data availability Statement

The data that support the findings of this study are available from the corresponding author upon reasonable request.

# Supplementary Material

# $a$-axis YBa$_2$Cu$_3$O$_{7-x}$/PrBa$_2$Cu$_3$O$_{7-x}$/YBa$_2$Cu$_3$O$_{7-x}$

# trilayers with subnanometer rms roughness

Y. Eren Suyolcu[1†], Jiaxin Sun[1†], Berit H. Goodge[2], Jisung Park[1], Jürgen Schubert[3], Lena F. Kourkoutis[2,4], and Darrell G. Schlom[1,4,5]

[1]Department of Materials Sciences and Engineering, Cornell University, Ithaca, New York 14853, USA

[2]School of Applied and Engineering Physics, Cornell University, Ithaca, New York 14853, USA

[3]Peter Grünberg Institute (PGI-9) and JARA-Fundamentals of Future Information Technology, Forschungszentrum Jülich GmbH, 52425 Jülich, Germany

[4]Kavli Institute at Cornell for Nanoscale Science, Ithaca, New York 14853, USA

[5]Leibniz-Institut für Kristallzüchtung, Max-Born-Str. 2, 12489 Berlin, Germany

[†]These authors contributed equally to this work

This file contains:

Supplementary Note, Table, and Supplementary Figures S1-S10

**Binary oxide calibration**

To achieve the desired Y(Pr):Ba:Cu = 1:2:3 composition ratio in the deposited Y(Pr)Ba$_2$Cu$_3$O$_{7-x}$ films, we assume that these elements have a sticking coefficient of unity for the growth conditions used for Y(Pr)Ba$_2$Cu$_3$O$_{7-x}$. Under this assumption, we establish the individual fluxes of yttrium, praseodymium, barium, and copper by synthesizing epitaxial films of their respective binary oxides individually, and use either x-ray reflectivity (XRR) or RHEED oscillations to determine the thicknesses of these calibration films. From the measured film thickness and assuming (1) unity sticking coefficients of these cations for the growth conditions used to grow the binary oxide calibration films and (2) that these calibration films are fully relaxed and have the bulk density of these binary oxides, we calculate the respective elemental fluxes. Having established the elemental fluxes under the assumptions stated, we then alter the temperatures of the MBE effusion cells containing yttrium, praseodymium, barium, and copper until the binary oxide calibration films grown with these sources indicate that we have the desired 1:2:3 flux ratio among Y(Pr):Ba:Cu. At this point, the growths of Y(Pr)Ba$_2$Cu$_3$O$_{7-x}$ and YBa$_2$Cu$_3$O$_{7-x}$/PrBa$_2$Cu$_3$O$_{7-x}$/YBa$_2$Cu$_3$O$_{7-x}$ trilayers commence. We perform these binary oxide flux calibrations each day, immediately prior to the growth of Y(Pr)Ba$_2$Cu$_3$O$_{7-x}$ and YBa$_2$Cu$_3$O$_{7-x}$/PrBa$_2$Cu$_3$O$_{7-x}$/YBa$_2$Cu$_3$O$_{7-x}$ trilayers to achieve the growth of stoichiometric films. Our demonstrated success in growing phase-pure Y(Pr)Ba$_2$Cu$_3$O$_{7-x}$ thin films and YBa$_2$Cu$_3$O$_{7-x}$/PrBa$_2$Cu$_3$O$_{7-x}$/YBa$_2$Cu$_3$O$_{7-x}$ trilayers attests to the validity of the assumptions we have made in this calibration procedure.

The conditions used to grow each of the binary oxides—Y$_2$O$_3$, PrO$_2$, BaO, and CuO—are outlined in Supplementary Table S1. Also shown are the substrates used and epitaxial orientations of the resulting binary oxide calibration films. In all cases, the oxidant used is distilled ozone (~80% $O_3$ + 20% $O_2$), i.e., the same oxidant and background pressure used for the growth of the subsequent $Y(Pr)Ba_2Cu_3O_{7-x}$ thin films and $YBa_2Cu_3O_{7-x}/PrBa_2Cu_3O_{7-x}/YBa_2Cu_3O_{7-x}$ trilayers. For $Y_2O_3$, $PrO_2$, and CuO the thickness of the calibration film is measured by XRR. For hydroscopic BaO, the RHEED oscillation periodicity corresponding to the smallest charge-neutral formula unit of BaO,[1] which in the case of BaO is half of the cubic lattice constant of BaO, is used to calculate the barium flux. Examples of the RHEED patterns, RHEED oscillations, XRD $\theta$-$2\theta$ scans, and XRR scans for this calibration method are shown for each binary oxide in Figs. S1-S4. The orientation relationships of the binary oxide calibration layers are: (111) $Y_2O_3$ ∥ (111) YSZ and $[\bar{1}10]$ $Y_2O_3$ ∥ $[\bar{1}10]$ YSZ,[2] (111) $PrO_2$ ∥ (111) YSZ and $[\bar{1}10]$ $PrO_2$ ∥ $[\bar{1}10]$ YSZ, (100) BaO ∥ (100) $SrTiO_3$ and [100] BaO ∥ [110] $SrTiO_3$,[3] and (111) CuO ∥ (100) MgO and $[\bar{1}10]$ CuO ∥ [110] MgO.[4]

**TABLE S1**. Binary oxide growth conditions, substrates used, and the orientation of the resulting epitaxial films.

| Binary oxide | Substrate | Temperature (°C) | Pressure (Torr) |
|---|---|---|---|
| (111) CuO | (100) MgO | Room Temperature | $1 \times 10^{-6}$ |
| (100) BaO | (100) $SrTiO_3$ | 600 | $1 \times 10^{-6}$ |
| (111) $Y_2O_3$ | (111) YSZ | 900 | $1 \times 10^{-6}$ |
| (111) $PrO_2$ | (111) YSZ | 900 | $1 \times 10^{-6}$ |

YSZ = $(ZrO_2)_{0.905}(Y_2O_3)_{0.095}$ (or 9.5 mol% yttria-stabilized zirconia)

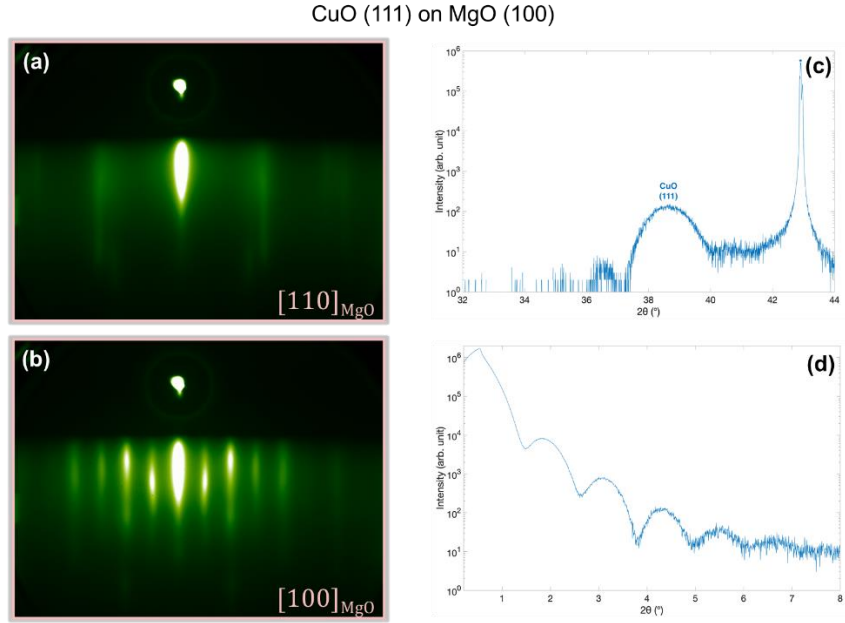

**Figure S1.** Growth of a CuO calibration film. (a) and (b) RHEED patterns along the [110] and [100] azimuths of MgO. (c) $\theta$-$2\theta$ XRD scan and (d) x-ray reflectivity scan. The example shown corresponds to a 12.2 nm thick CuO calibration film.

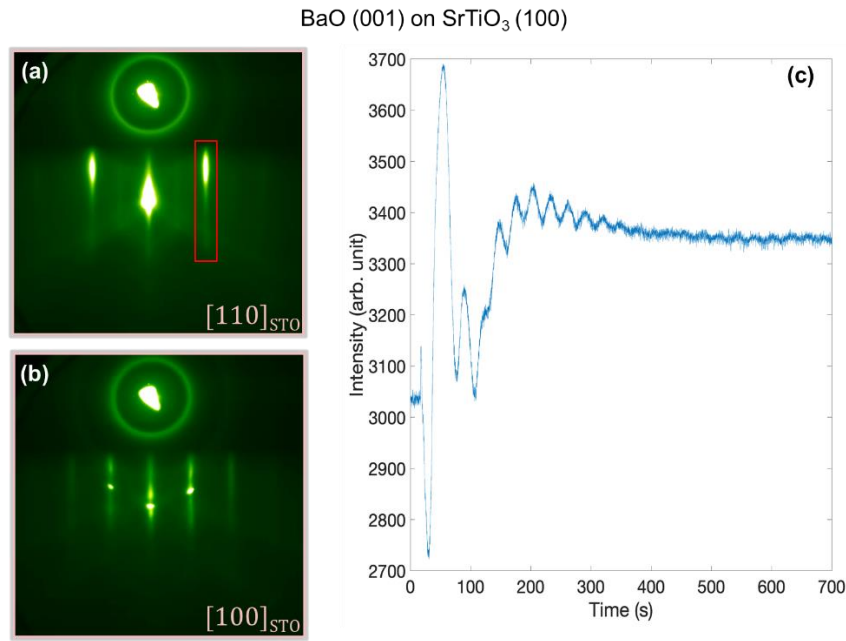

**Figure S2.** Growth of a BaO calibration film. (a) and (b) RHEED patterns along the [110] and [100] azimuths of $SrTiO_3$ and (c) RHEED oscillations of the diffracted streak from the outlined red box showing layer-by-layer growth. The example shown corresponds to a ~7 nm thick BaO calibration film.

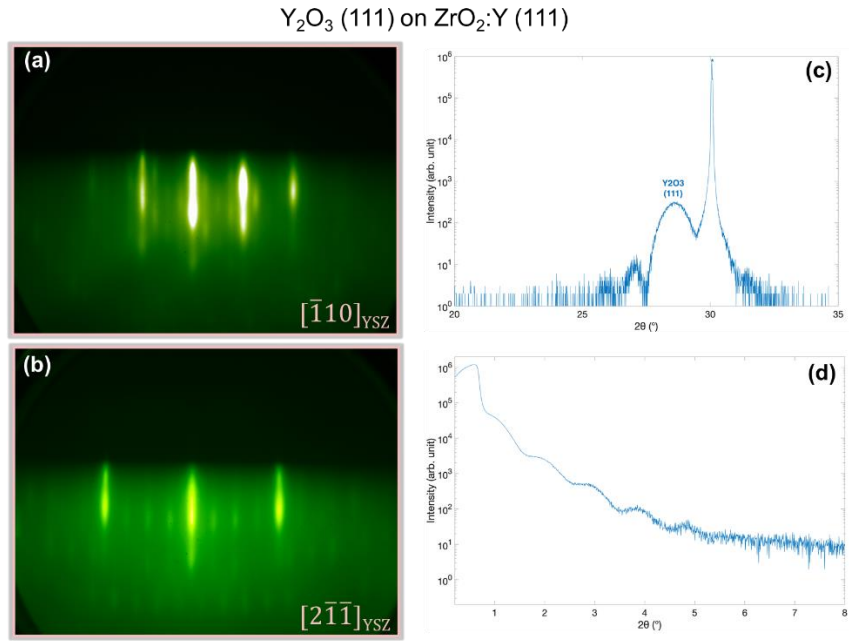

**Figure S3.** Growth of a $Y_2O_3$ calibration film. (a) and (b) RHEED patterns along the $[\bar{1}10]$ and $[2\bar{1}\bar{1}]$ azimuths of YSZ. (c) $\theta$-$2\theta$ scan and (d) x-ray reflectivity scan. The example shown corresponds to a 11.5 nm thick $Y_2O_3$ calibration film.

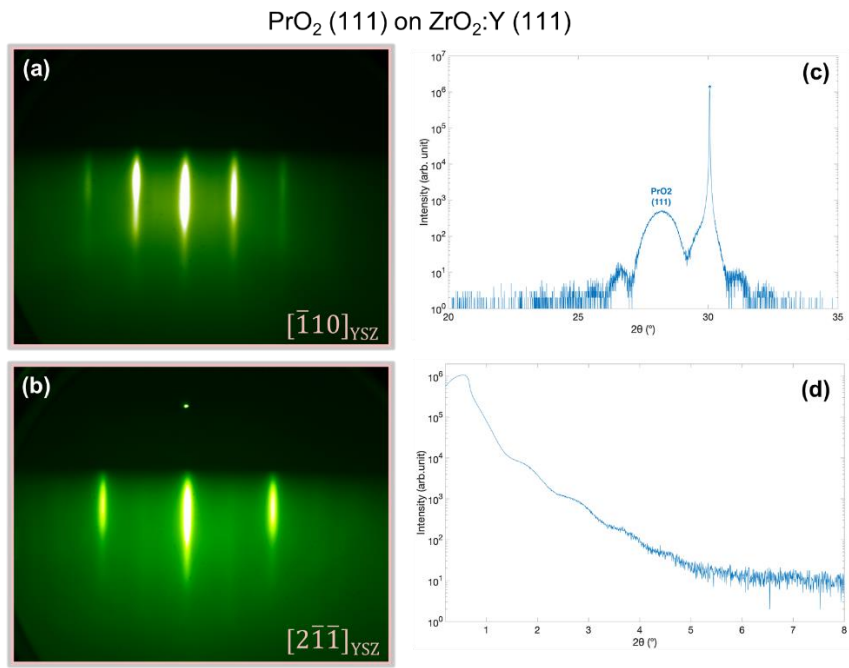

**Figure S4.** Growth of a $PrO_2$ calibration film. (a) and (b) RHEED patterns along the $[\bar{1}10]$ and $[2\bar{1}\bar{1}]$ azimuths of YSZ. (c) $\theta$-$2\theta$ scan and (d) x-ray reflectivity scan. The example shown corresponds to a 11.1 nm thick $PrO_2$ calibration film.

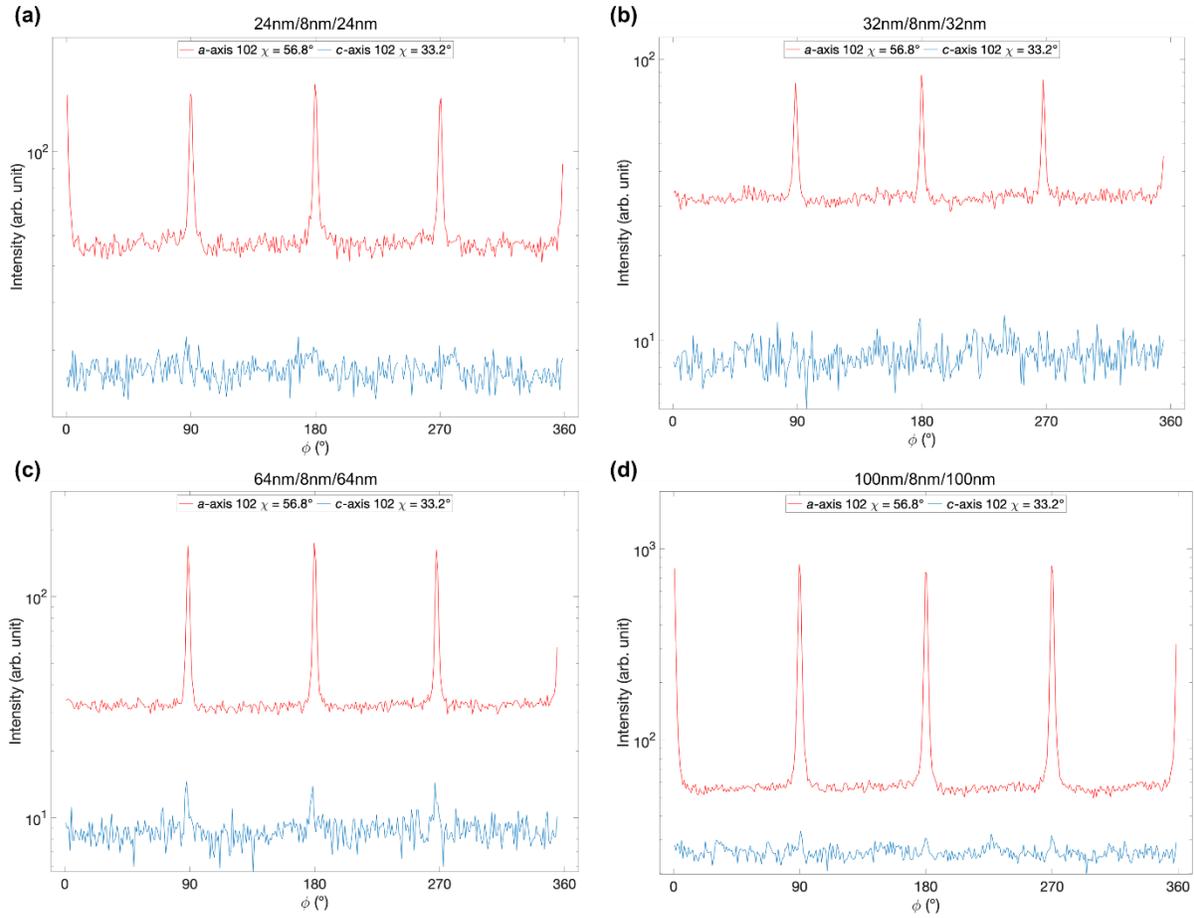

**Figure S5.** Off-axis $\phi$-scans of the 102 family of peaks at $\chi \approx 56.8°$ (red) and $\chi \approx 33.2°$ (blue) of (a) 24 nm/8 nm/24 nm, (b) 32 nm/8 nm/32 nm, (c) 64 nm/8 nm/64 nm, and (d) 100 nm/8 nm/100 nm YBa$_2$Cu$_3$O$_{7-x}$/PrBa$_2$Cu$_3$O$_{7-x}$/YBa$_2$Cu$_3$O$_{7-x}$ trilayers. Peak area integration indicates that the volume fractions of $a$-axis oriented Y(Pr)Ba$_2$Cu$_3$O$_{7-x}$ in these trilayers are 97%, 97%, 96%, and 99%, , respectively.

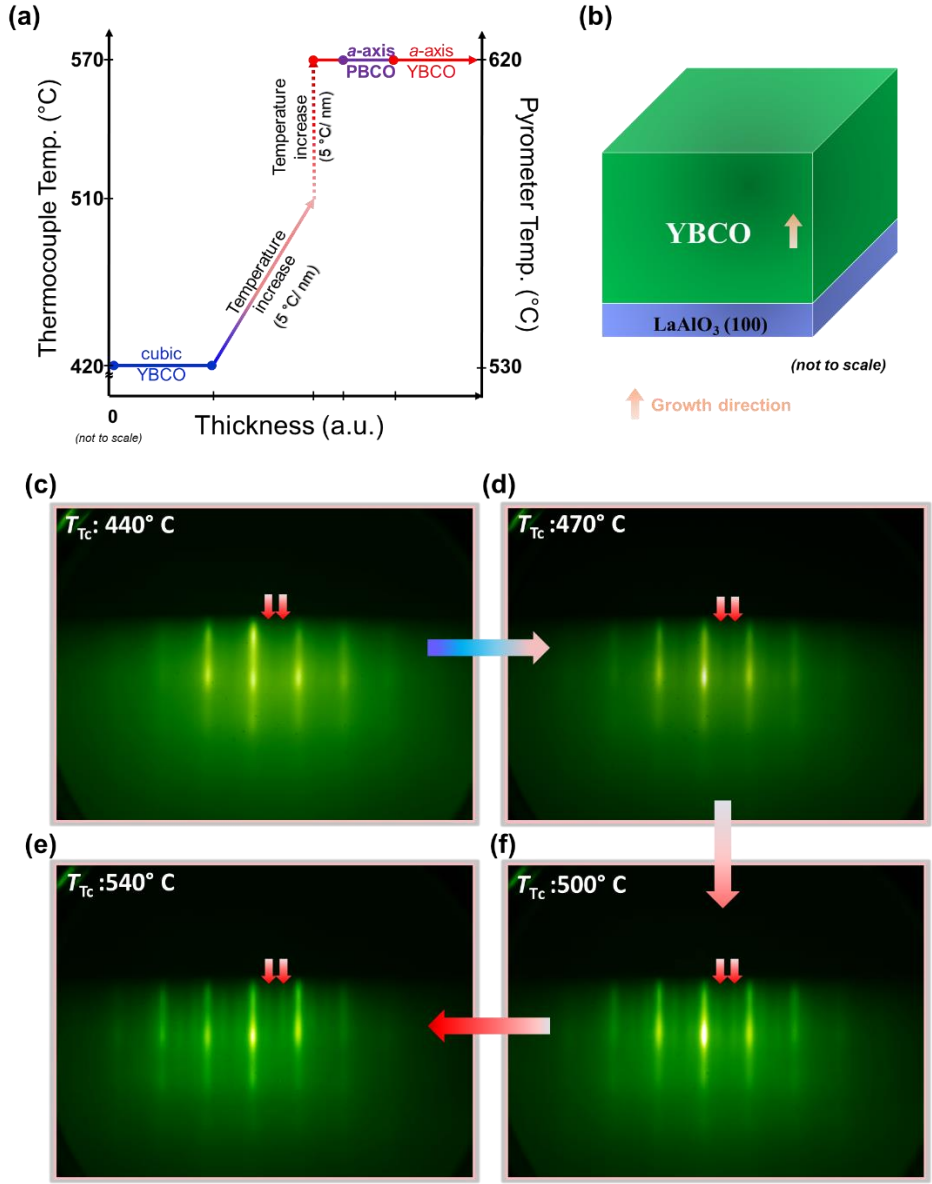

**Figure S6.** Temperature ramping procedure and related RHEED patterns. (a) A simple depiction of the four-step temperature-ramping method for the growth of the $YBa_2Cu_3O_{7-x}$/$PrBa_2Cu_3O_{7-x}$/$YBa_2Cu_3O_{7-x}$ trilayers. (b) Schematic of the $YBa_2Cu_3O_{7-x}$ layer grown on a (100) $LaAlO_3$ substrate. Real-time RHEED images of a single $YBa_2Cu_3O_{7-x}$ layer acquired at four different temperatures during the temperature ramping procedure, i.e., at thermocouple temperatures of (c) $T_{Tc} = 440$ °C, (d) $T_{Tc} = 470$ °C, (e) $T_{Tc} = 500$ °C, and (f) $T_{Tc} = 540$ °C. The red arrows added to (c)-(f) point to the diffraction streaks associated with the $c$-axis of the $YBa_2Cu_3O_{7-x}$ lying in-plane. As the temperature is ramped, these streaks become more intense and sharper, demonstrating the enhanced crystalline quality of the $a$-axis oriented $YBa_2Cu_3O_{7-x}$ film during growth.

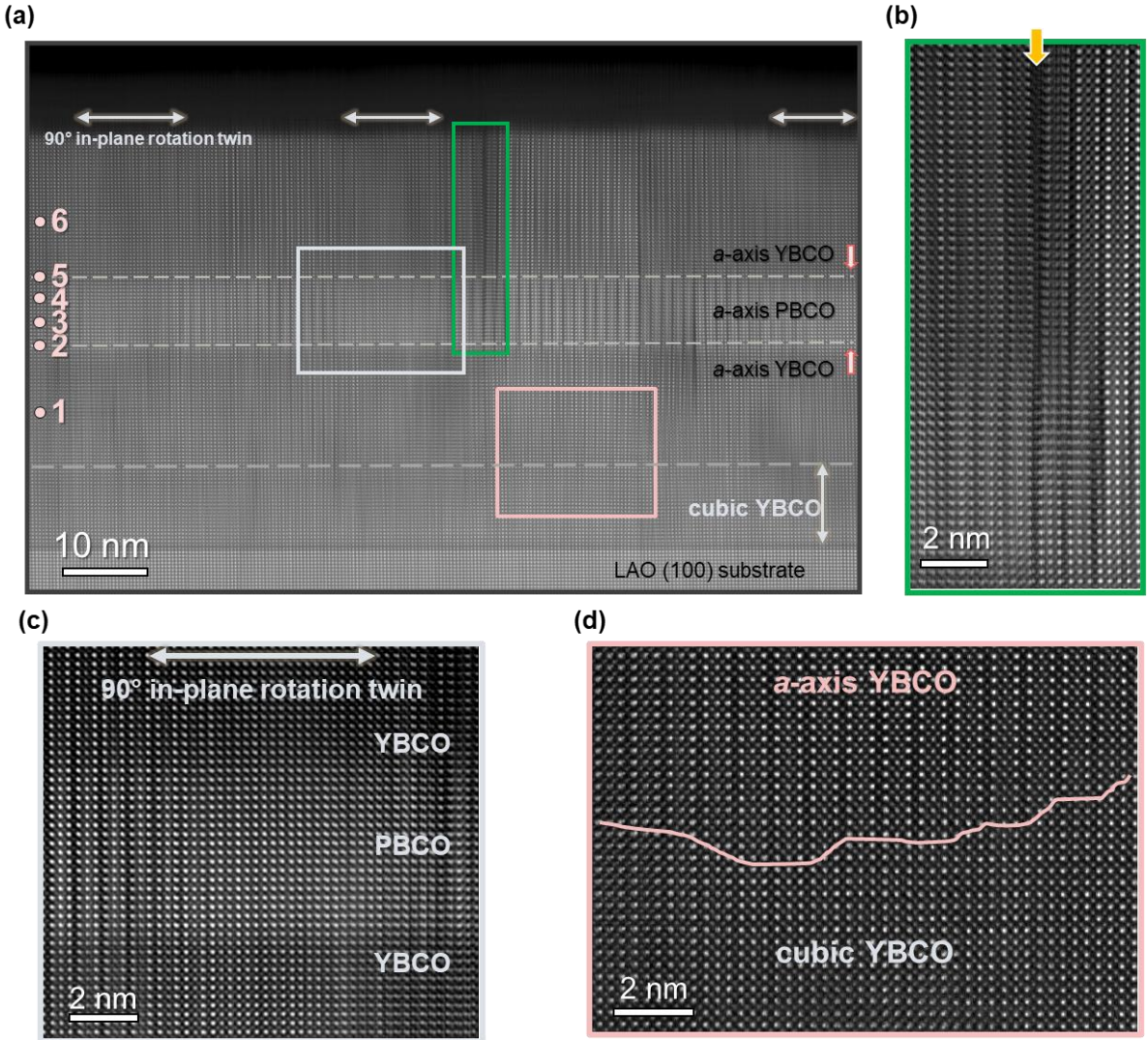

**Figure S7.** (a) Representative cross-sectional HAADF-STEM image of the 24 nm/8 nm/24 nm $YBa_2Cu_3O_{7-x}$/$PrBa_2Cu_3O_{7-x}$/$YBa_2Cu_3O_{7-x}$ trilayer highlighting aspects of the microstructure observed. The six (1-6) pink circles indicate the positions at which the RHEED patterns presented in Fig. 1(a) in the main text were taken during growth. The thickness of the cubic $(Y,Ba)CuO_{3-x}$ buffer layer is 10±2 nm. (b) High-magnification HAADF-STEM image showing the region of interest highlighted by the green rectangle in (a). The orange arrow indicates the intercalation of an additional CuO layer (dark feature) as a consequence of $YBa_2Cu_4O_{8-x}$ stacking faults with double-chain layers. (c) The high magnification view of the area highlighted by the gray rectangle in (a) demonstrates the presence of $YBa_2Cu_3O_{7-x}$ and $PrBa_2Cu_3O_{7-x}$ in-plane rotation twins on top of the cubic $(Y,Ba)CuO_{3-x}$ layer that forms during the "cold" growth. The gray double-sided arrows mark the width of the 90° in-plane rotational twins. (d) High magnification view of the area highlighted by the pink rectangle in (a) shows the boundary between the cubic $(Y,Ba)CuO_{3-x}$ and the *a*-axis $YBa_2Cu_3O_{7-x}$ layer, marked by the pink line.

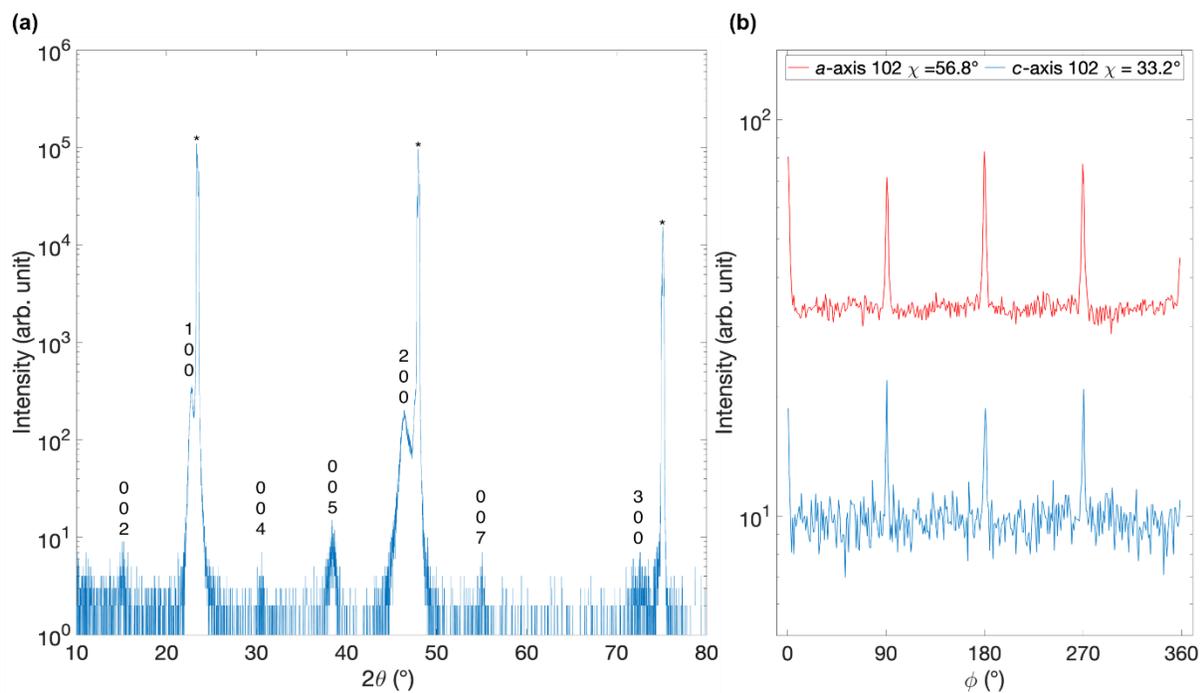

**Figure S8.** XRD scans of the less-ideal 32 nm/8 nm/32 nm $YBa_2Cu_3O_{7-x}$/$PrBa_2Cu_3O_{7-x}$/$YBa_2Cu_3O_{7-x}$ trilayer containing a higher concentration of $c$-axis grains. (a) $\theta$-$2\theta$ scan and (b) $\phi$-scan of the 102 family of peaks at $\chi \approx 56.8°$ (red) and $\chi \approx 33.2°$ (blue) of this less ideal trilayer indicate that it contains a volume fraction of ~84% $a$-axis $YBa_2Cu_3O_{7-x}$.

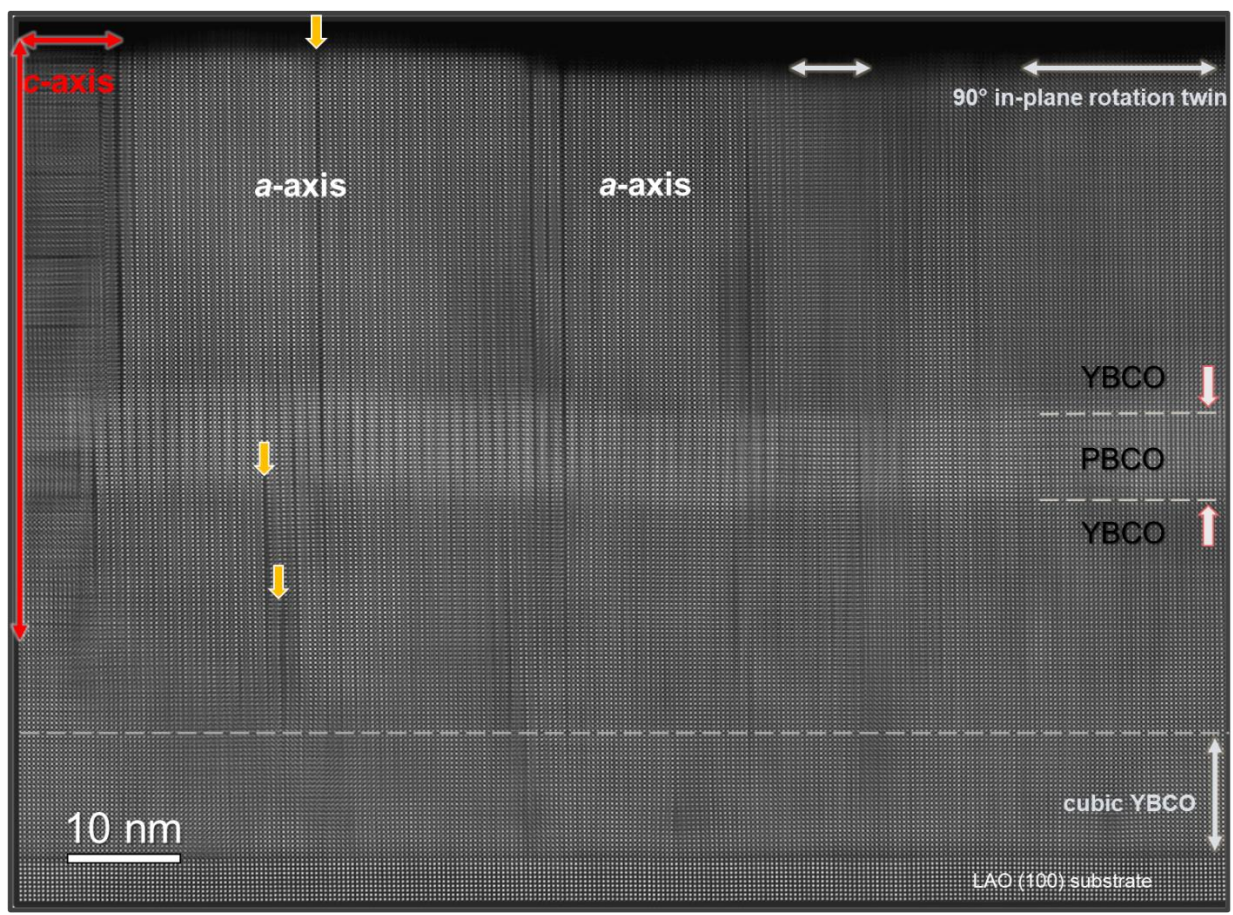

**Figure S9.** (a) Representative cross-sectional HAADF-STEM image of the less-ideal 32 nm/8 nm/32 nm YBa$_2$Cu$_3$O$_{7-x}$/PrBa$_2$Cu$_3$O$_{7-x}$/YBa$_2$Cu$_3$O$_{7-x}$ trilayer showing increased roughness, $c$-axis and $a$-axis domains as well as YBa$_2$Cu$_4$O$_{8-x}$ stacking faults (dark lines), and a cubic (Y,Ba)CuO$_{3-x}$ buffer layer. The red arrows mark the $c$-axis domain that starts from the bottom cubic (Y,Ba)CuO$_{3-x}$ layer. The orange arrows indicate some of the YBa$_2$Cu$_4$O$_{8-x}$ stacking faults (features with darker contrast). The gray double-ended arrows mark the width of examples of 90° in-plane rotational twins in which the $c$-axis is oriented into the plane of the image. The thickness of the cubic-(Y,Ba)CuO$_{3-x}$ buffer layer is 11±2 nm.

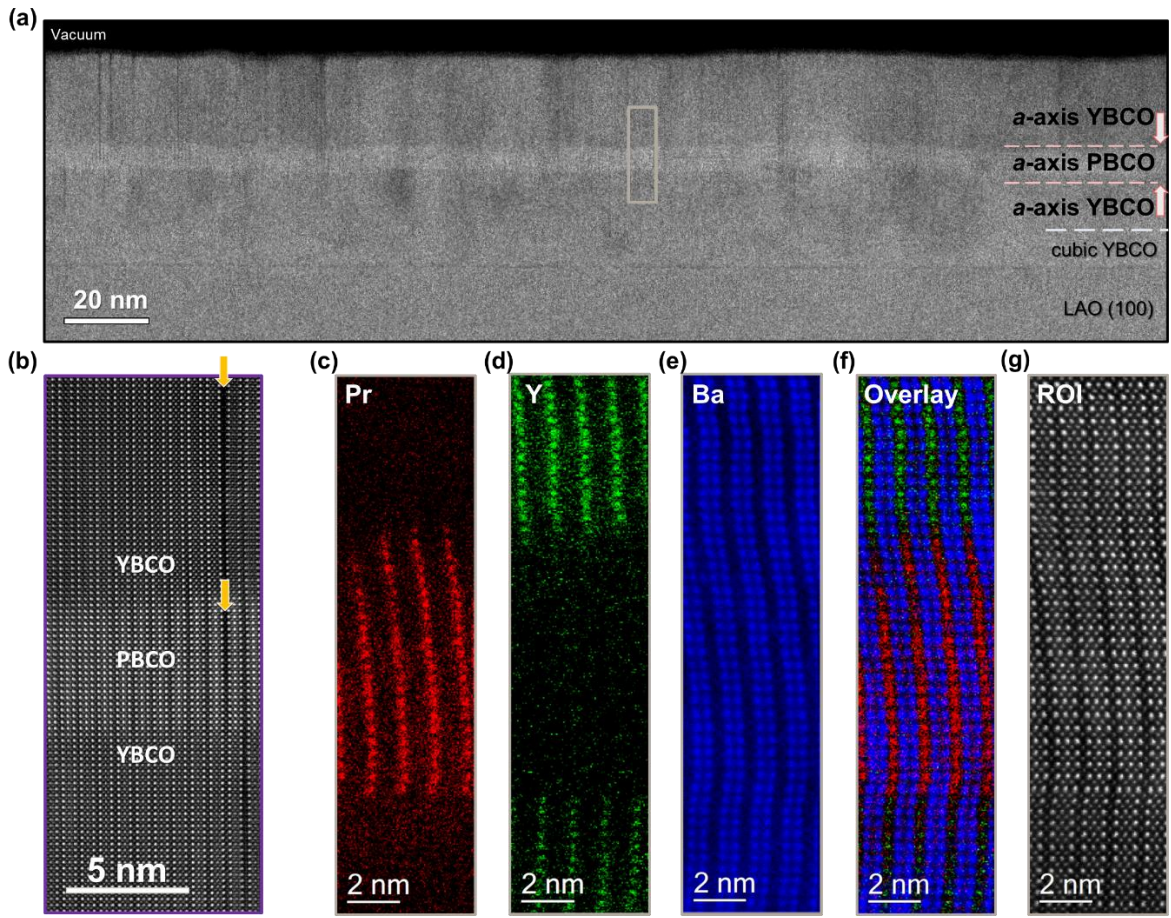

**Figure S10.** (a) Low-magnification cross-sectional HAADF-STEM image of the same less-ideal 32 nm/8 nm/32 nm YBa$_2$Cu$_3$O$_{7-x}$/PrBa$_2$Cu$_3$O$_{7-x}$/YBa$_2$Cu$_3$O$_{7-x}$ trilayer exhibiting increased roughness. Individual YBa$_2$Cu$_3$O$_{7-x}$ and PrBa$_2$Cu$_3$O$_{7-x}$ layers are separated using dashed lines and the pink arrows indicate the nominal interfaces. (b) The high-magnification scan of a representative region demonstrates fairly abrupt interfaces and YBa$_2$Cu$_4$O$_{8-x}$ stacking faults as marked with orange arrows that either originate from the underlying YBa$_2$Cu$_3$O$_{7-x}$ or PrBa$_2$Cu$_3$O$_{7-x}$ layers. (c–e) Atomically resolved Pr–$M_{5,4}$ edge (red), Y–$L_{3,2}$ edge (green), and Ba–$M_{5,4}$ edge (blue) elemental maps evidencing the sharp chemical abruptness of the interfaces. (f) The RGB overlay and (g) the simultaneously acquired ADF-STEM image showing this same region highlighted by the tan rectangle in (a).